\documentclass{llncs}
\usepackage{logic9}
\usepackage{latexsym}
\usepackage{amssymb,amsmath,amsfonts}
\usepackage{tikz}
\usetikzlibrary{automata}
\usetikzlibrary[arrows]

\newcommand {\bg}{\mathit{beg}}

\newcommand {\en}{\mathit{end}}

\newcommand{\rel}{\mathit{rel}}
\newcommand{\acq}{\mathit{acq}}

\newcommand{\opD}{\mathrel{D}}
\newcommand{\loc}{\mathit{dom}}

\newcommand{\init}{\mathit{init}}
\newcommand{\out}{\mathit{out}}

\title{Automated Synthesis of Distributed Controllers}

\author{Anca Muscholl}
\institute{LaBRI, University of Bordeaux}

\begin{document}

\maketitle

%\section{Introduction}

% \begin{itemize}
% \item testing vs verifying concurrent programs; even better:
%   monitoring/exception handling
% \item interleaving-based approaches: too unreliable for testing and
%   too expensive for verification; solution: context
%   switches (CHESS)
% \item POR and stateless exploration (Godefroid, Parosh)
% \end{itemize}

\begin{abstract}
  Synthesis is a particularly challenging problem for concurrent programs.
  At the same time it is a very promising approach, since concurrent programs
  are difficult to get right, or to analyze with
  traditional verification techniques. This paper gives an introduction to
  distributed synthesis in the setting of Mazurkiewicz traces, and its
  applications to decentralized runtime monitoring.
\end{abstract}

\section{Context}

Modern computing systems are increasingly distributed and
heterogeneous. Software needs to be able to exploit these advances,
providing means for applications to be more performant.  Traditional concurrent
programming paradigms, as in Java, are based on threads,
shared-memory, and locking mechanisms that guard access to common
data. More recent paradigms like the reactive programming model
of Erlang~\cite{erlang07} and Scala~\cite{scala14,scala-book} replace shared
memory by asynchronous message 
passing, where sending a message is non-blocking.

In all these concurrent frameworks, writing reliable software is a
serious challenge. Programmers tend to think about code mostly in a
sequential way, and it is hard to grasp all possible schedulings
of events in a concurrent execution. For similar reasons, verification and analysis
of concurrent programs is a difficult task. Testing, which is still
the main method for error detection in software, has low coverage for
concurrent programs. The reason is that bugs in such programs are
difficult to reproduce: they may happen under very specific thread
schedules and the likelihood of taking such corner-case schedules is
very low. Automated verification, such as model-checking and other
traditional exploration techniques, can handle very limited instances
of concurrent programs, mostly because of the very large number of
possible states and of possible interleavings of executions.

% The second strategy consist in exploring a well-chosen subset of
% interleaved executions, for instance only interleavings with a bounded
% number of \emph{context switches} between threads~\cite{}. For many
% classes of programs, experiments have shown that bugs are typically
% discovered with small bounds on context switches.

%\section{Concurrent programs}

Formal analysis of programs requires as a pre-requisite a
clean mathematical model for programs. Verification of
sequential programs starts usually with an abstraction step -- reducing the value
domains of variables to finite domains, viewing conditional branching
as non-determinism, etc. Another major simplification consists in
disallowing recursion. This leads to a very robust computational
model, namely \emph{finite-state automata} and \emph{regular
  languages}. Regular languages of words (and trees) are particularly
well understood notions. The deep connections between logic and
automata revealed by the foundational work of B\"uchi, Rabin, and
others, are the main ingredients in automata-based verification.

In program synthesis, the task is to turn a specification into a
program that is guaranteed to satisfy it. Synthesis can therefore provide
solutions that are correct by construction. It is thus
particularly attractive for designing concurrent programs, that are
often difficult to get right or to analyze by traditional
methods. In distributed synthesis, we are given in addition an
architecture, and the task is to turn the specification into a distributed
implementation over this architecture. 

Distributed synthesis proves to be a real challenge, and there are at
least two  reasons for this. First, there is no canonical model for concurrent
systems, simply because there are very different kinds of
interactions between processes. Compare, for example, multi-threaded
shared memory Java programs, to Erlang or Scala programs with
asynchronous function calls. This issue is connected with another, more
fundamental reason: techniques for distributed synthesis are rather
rare, and decidability results are conditioned by the right match
between the given concurrency model and the kind of questions  that we ask.

\emph{Mazurkiewicz traces} were introduced in the late seventies by
A.~Mazurkiewicz \cite{maz77} as a simple model for concurrency
inspired by Petri nets. Within this theory, Zielonka's
theorem~\cite{zie87} is a prime example of a result on distributed
synthesis. 

This paper gives a brief introduction to Mazurkiewicz
traces and to Zielonka's theorem, and describes how this theory can be
used in the verification and the design of concurrent programs. We
focus on the synthesis of concurrent programs and its application to
decentralized runtime monitoring. 

Monitoring is a more lightweight alternative to model-checking and
synthesis. The task is to observe the execution of a program  in order to detect
possible violations of safety requirements. Monitoring is a
prerequisite for control, because it can gather information about
things that went wrong and about components that require repair
actions. In programming, monitoring takes the form of assertions: an
invalidation of an assertion is the first sign that something has gone
wrong in the system. However, concurrent programs often require
assertions concerning several components. A straightforward but
impractical way to verify such an assertion at runtime is to
synchronize the concerned components and to inquire about their
states. A much better way to do this is to write a distributed monitor
that deduces the required information  by
recording and exchanging suitable information using the available
communication in the program. Mazurkiewicz trace
theory and Zielonka's theorem can provide a general, and yet practical
method for synthesizing distributed monitors.

\medskip

\emph{Overview of the paper.} Section~\ref{sec:intro} sets the stage
by describing some classical correctness issues for concurrent
programs. Section~\ref{sec:traces} introduces 
Mazurkiewicz traces, and Section~\ref{sec:moni} presents 
some applications to decentralized monitoring.

\section{Distributed models: some motivation}\label{sec:intro}

Concurrent programming models usually consist of entities, like
processes or threads, that evolve in an asynchronous manner and
synchronize on joint events, such as access to shared variables, or
communication.  We start with some illustrating examples from
multi-threaded programming, and with some typical correctness
properties. This will allow us to present the type of questions that
we want to address.

A multi-threaded program consists of an arbitrary number of
concurrently executing threads. We will assume that there is a fixed
set $\Tt$ of threads.  There is no global clock, so threads progress
asynchronously. Threads can either perform local actions or access the
global memory, consisting of shared variables from a fixed set
$X$. Possible actions of a thread $T \in \Tt$ include reads $r(T,x,v)$
and writes $w(T,x,v)$ on a shared variable $x \in X$ (for some value
$v$) and acquiring $\acq(T,L)$, resp.~releasing $\rel(T,L)$ a lock
$L$. More complex forms of access to the shared memory, such as
compare-and-set (CAS), are commonly used in lock-free programming. We
will not use CAS in the remaining of this section, but come back to it in
Section~\ref{sec:traces}.

% \begin{example}
%   The CAS operation is available as atomic operation in the
%   java.util.concurrent.atomic package, and supported by many
%   architectures. It takes as parameters the thread identifier $T$, the
%   variable name $x$, and two values, $v_\mathit{old}$ and
%   $v_\mathit{new}$. The effect of CAS($T,x,v_\mathit{old},
%   v_\mathit{new})$ is conditional: the value of $x$ is replaced by
%   $v_\mathit{new}$ if it is equal to $v_\mathit{old}$, otherwise it
%   does not change. The method returns true if the value was swapped,
%   and false oterhwise.
% \end{example}

Partial orders are a classical abstraction for reasoning about
executions of multi-threaded programs. The computation on each thread
is abstracted out by a set of events, and the multi-threaded execution
is abstracted in form of a partial order on these events.  An early
example is Lamport's \emph{happens-before} relation~\cite{lamport78}, originally
described for communicating systems. This relation orders the events on each thread, and the
sending of a message before its receive. In multi-threaded programs
with shared memory, where locks guard the access to shared
variables, the happens-before relation orders two events if they are performed by
the same thread or they use the same lock.

A more formal, general definition of the happens-before relation for
programs goes as
follows. Let $\S$ be the set of actions in a program. We will assume
throughout the paper that $\S$ is finite. Depending on the problem
that we consider, we will assume that there is a binary
\emph{conflict} relation $D \subseteq \S \times \S$ between the
actions of the program. For example, we will have $a \opD b$ if $a$
and $b$ are performed by the same thread. Starting with a linear
execution $a_1 \cdots a_n \in \S^*$ of a program, the~\emph{happens-before}
relation is the partial order on positions defined as the
reflexive-transitive closure of the relation $\set{i \prec j \mid i<j
  \text{ and } a_i \opD a_j}$. As we will see in
Section~\ref{sec:traces}, if the conflict relation is symmetric, this
partial order is a \emph{Mazurkiewicz
  trace}.

In the remaining of this section we outline two frequently considered
correctness issues for concurrent programs, that will be used as
examples for decentralized monitoring in Section~\ref{sec:moni}.

\subsection{Race detection}\label{ssec:races}

Race detection is one of the widely studied problems of concurrent
software. Informally, a race occurs whenever there are conflicting
accesses to the same variable without proper
synchronization. Detecting races is important since executions with
races may yield unexpected behaviors, caused by the outcome of the computation
depending on the schedule of threads.

In order to define races for multi-threaded programs with
lock synchronization we need to introduce the happens-before relation
for such programs. Let $\S$ be the set of actions in a program,
for instance:
\[
\S=\set{w(T,x), r(T,x), \acq(T,L), \rel(T,L) \mid T,x,L}\,.
\]
Two actions from $\S$ are in \emph{conflict}  \label{page:D} if
\begin{itemize}
\item they are performed by the same thread, or
\item they acquire or release the same lock.
\end{itemize}
%We write $a \opD b$ if $a,b$ are in conflict.   

% Two accesses to the same variable are called conflicting, if at
% least one of them is a write. 

A \emph{race} occurs in an execution if there are two 
accesses to the same shared variable such that

\begin{itemize}
\item they are \emph{unordered} in the happens-before relation, and
\item at least one of them is a write.
\end{itemize}

\begin{example}
  Figure~\ref{fig:race} illustrates a race problem due to locking that
  is too fine-grained. Two threads have access to a list pointed to by
  \texttt{head}. \texttt{Thread 1} adds an element to the head of the
  list, while \texttt{Thread 2} deletes the head element. The two
  instructions protected by the lock are ordered in the happens-before
  relation. However, \texttt{t1.next = head} and \texttt{head =
    head.next} are unordered. Since the first instruction is a read,
  and the second a write, this situation is a race condition.

  \begin{figure}
    \centering
    \begin{minipage}{.49\linewidth}
 \begin{verbatim}
type list {int data; list *next}
list *head

Thread 1

1: t1 = new(list);
2: t1.data = 42;
3: t1.next = head;
4: ack(lock)
5:   head = t1
6: rel(lock)
\end{verbatim}
  \end{minipage}
  \begin{minipage}{.49\linewidth}
\begin{verbatim}
Thread 2

7:  t2 = head;
8:  ack(lock)
9:   head = head.next
10: rel(lock)
\end{verbatim}
  \end{minipage}
    \caption{A race condition in the execution $1,2,3,7,8,9$: events
     $3$ and $9$ are unordered in the happens-before relation.}
    \label{fig:race}
  \end{figure}
  
\end{example}

\subsection{Atomicity}\label{ssec:atomic}

Atomicity, or conflict serializability, is a high-level correctness
notion for concurrent programs that has its origins in database
transactions. A transaction consists of a block of operations, such as
reads and writes to shared memory variables, that is marked as
\texttt{atomic}. An execution of the transaction system is serial if
transactions are scheduled one after the other, without interleaving
them.  A serial execution reflects the intuition of the programmer,
about parts of the code marked as  transactions as being executed
atomically. 

In order to define when a multi-threaded program is
conflict-serializable, we need first the notion of equivalent
executions. Two executions are \emph{equivalent} if they define the
same happens-before relation w.r.t.~the following conflict relation:
Two actions from $\S=\set{w(T,x),r(T,x) \mid T, x}$ are in conflict if
\begin{itemize}
\item they are performed by the same thread, or
\item they access to the same variable, and at least one of them is a
  write.
\end{itemize}

The above conflict relation  has a different purpose than the one
used for the race problem: here, we are interested in the values
that threads compute. Two executions are considered to be equivalent if
all threads end up with the same values of (local and global)
variables. Since a write $w(T,x)$ potentially modifies the value of
$x$, its order w.r.t.~any other access to $x$ should be
preserved. This guarantees that the values of $x$ are the same in
two equivalent executions.

A program is called \emph{conflict-serializable} (or \emph{atomic}) if
every execution is equivalent to a serial one.  As we will explain
in Section~\ref{sec:traces} this means that every
execution can be reordered into an equivalent one where no transaction is
interrupted.

\begin{example}\label{ex:atomic}
  Figure~\ref{fig:atomic} shows a simple program with two
    threads that is not conflict-serializable. The
  interleaved execution where \texttt{Thread 2} writes after the read
  and before the write of \texttt{Thread 1}, is not equivalent to any
  serial execution.

  \begin{figure}
    \centering
    \begin{minipage}{.49\linewidth}
 \begin{verbatim}
Thread 1

1: atomic {
2:   read(x);
3:   write(x)
4:   }
\end{verbatim}
  \end{minipage}
  \begin{minipage}{.49\linewidth}
\begin{verbatim}
Thread 2

5: atomic {
6:   write(x);
7:   }
\end{verbatim}
  \end{minipage}
    \caption{A program that is not conflict-serializable: the
      execution $1,2,5,6,7,3,4$ is not equivalent to any serial execution.}
    \label{fig:atomic}
  \end{figure}

\end{example}

\section{Mazurkiewicz traces and Zielonka's theorem}\label{sec:traces}

This section introduces Mazurkiewicz traces~\cite{maz77}, one of the
simplest formalisms able to describe concurrency. We will see that
traces are perfectly suited to describe dependency and the
happens-before relation. The notion of conflicting actions and
the happens-before relation seen in the previous section are
instances of this more abstract approach.

The definition  of traces starts with an alphabet of actions $\S$ and a
\emph{dependence relation} $D \subseteq \S \times \S$ on actions, that
is reflexive and symmetric. The idea behind this relation is that
two dependent actions are always ordered, for instance because the
outcome of one action affects the other action. For example, the
actions of acquiring or releasing the same lock are ordered, since a
thread has to wait for a lock to be released before acquiring it. 

\begin{example}\label{ex:D}
   Coming back to the problems introduced in
   Sections~\ref{ssec:races} and~\ref{ssec:atomic}, note that the conflict
   relations defined there are both symmetric. For example, we can define the
   dependence relation $D$ over the alphabet 
$\S=\set{r(T,x), w(T,x) \mid T \in \Tt, x \in X}$ of
Section~\ref{ssec:atomic}, by letting $a \opD b$ if $a,b$ are in conflict.
% \begin{itemize}
% \item $a,b$ are actions of the same thread $T$, or
% \item $a,b$ access to the same variable $x \in X$ and at least one of
%   them is a write.
% \end{itemize}
\end{example}

While the dependence relation
coincides with the conflict relation, the happens-before relation
is the~\emph{Mazurkiewicz trace} order. A Mazurkiewicz trace
is
the labelled partial order  $T(w)=\di{E,\preceq}$ obtained from a
word $w=a_1 \ldots a_n
\in \S^*$ in the following way:
\begin{itemize}
\item $E=\set{e_1,\ldots,e_n}$ is the set of \emph{events}, in
  one-to-one correspondence with the positions of $w$, where event $e_i$ has
  label $a_i$,
\item $\preceq$ is the reflexive-transitive closure of $\set{(e_i,e_j)
    \mid i<j,\; a_i \opD a_j}$.
\end{itemize}

From a language-theoretical viewpoint, traces are almost as
attractive as words, and several results from automata and logics
generalize from finite and infinite words  to traces,
see~e.g.~the
handbook~\cite{DR95}. One of the cornerstone results in Mazurkiewicz
trace theory is based on an elegant  notion of finite-state distributed automata,
\emph{Zielonka automata}, that we present in the remaining of the section.

Informally, a Zielonka automaton~\cite{zie87} is a finite-state
automaton with control distributed over several~\emph{processes} that
synchronize on shared actions.  Synchronization is modeled through a
distributed action alphabet. There is no global clock: for instance
between two synchronizations, two processes can do a different number
of actions. Because of this, Zielonka automata are also known as
\emph{asynchronous automata}.

A \emph{distributed action alphabet} on a finite set $\PP$
of processes is a pair $(\S,\loc)$, where $\S$ is a finite set of
\emph{actions} and $\loc:\S \to (2^{\PP}\setminus \es)$ is a
\emph{domain function}. The domain $\loc(b)$ of action $b$
comprises all processes that synchronize in order to perform $b$. The domain function induces a natural dependence relation $D$ over
$\S$ by setting $a \opD b$ if $\loc(a) \cap \loc(b) \not=\es$. The
idea behind is that executions of two dependent actions affect at
least one common process, so their order matters. By contrast, two
\emph{independent actions} $a,b$, i.e., where $\loc(a) \cap  \loc(b)=\es$,  can
be executed  either as $ab$ or as $ba$, the order is immaterial.

\begin{example}\label{ex:procs}
  We reconsider Example~\ref{ex:D}. The dependence relation $D$
  defined there can be realized by a distributed alphabet $(\S,\loc)$
  on the following set of processes: 
  \[\PP=\Tt \cup \set{\struct{T,x} \mid T \in \Tt, x \in
    X}\,.\] Informally, each thread $T \in\Tt$ represents a process; in addition,
  there is a process for each pair $\struct{T,x}$. The process
  $\struct{T,x}$ stands for the cached value of $x$ in thread $T$.

The domain function defined below satisfies $a \opD b$ iff
$\loc(a) \cap \loc(b) \not=\es$:
\[
\loc(a) = 
\begin{cases}
  \set{T,\struct{T,x}} & \text{if } a=r(T,x)\\
\set{T,\struct{T',x} \mid T' \in \Tt} & \text{if } a =w(T,x)\, .
\end{cases}
\] 
The intuition behind $\loc(a)$ is as follows. A read $r(T,x)$ depends
both on the internal state of thread $T$ and the cached value of $x$,
and will affect the state of $T$. A write $w(T,x)$ depends on the
internal state of thread $T$ and will affect not only the state of
$T$, but also the cached values of $x$ on other threads using $x$,
since the new value will be written into these caches.
\end{example}

\noindent
A~\emph{Zielonka automaton} $\Aa=\struct{(S_p)_{p\in
    \PP},\; (s^\init_p)_{p\in\PP},\; \d}$ over $(\S,\loc)$ consists of:
\begin{itemize}
\item a finite set $S_p$ of (local) states with an initial state
  $s^\init_p \in S_p$,  for every process $p \in \PP$,
\item a transition relation
  $\d \subseteq \bigcup_{a \in\S} \big ( \prod_{p\in \loc(a)}S_p \times
  \set{a} \times \prod_{p\in
    \loc(a)}S_p \big )$. %  on tuples of states of processes in $\loc(a)$.
\end{itemize}

For convenience, we abbreviate a tuple $(s_p)_{p \in P}$ of local
states by $s_P$. An automaton is called \emph{deterministic} if the
transition relation is a partial function. 

Before explaining the semantics of Zielonka automata, let us
comment the idea behind the transitions and illustrate it through an
example. The reader may be more
familiar with synchronous products of finite automata, where a joint
action means that every automaton having this action in its alphabet
executes it according to its transition relation. Joint transitions in
Zielonka automata follow a \emph{rendez-vous} paradigm, meaning that
processes having action $b$ in their alphabet can exchange
information via the execution of $b$: a transition on $b$ depends on
the states of all processes executing $b$. The following example
illustrates this effect:

\begin{example}\label{ex:cas}
 %  Boolean multi-threaded programs with shared
%   variables can be modelled as Zielonka automata. As an example we
%   describe the translation for the \emph{compare-and-swap} (CAS)
%   instruction. This instruction has $3$ parameters: \textsf{CAS}($x$:
%   \textsf{variable}; \emph{old}, \emph{new}: \textsf{int}). Its effect is to return the
%   value of $x$ and at the same time set the value of $x$ to
%   \emph{new}, but only if the previous value of $x$ was equal to
%   \emph{old}. The compare-and-swap operation is a  widely used primitive
%   in implementations of concurrent data structures, and has
%   hardware support in most contemporary multiprocessor
%   architectures.
 The \texttt{CAS} (compare-and-swap) operation is available as
  atomic operation in the JAVA package java.util.concurrent.atomic, and
  supported by many architectures. It takes as parameters the thread
  identifier $T$, the variable name $x$, and two values,
  \texttt{old} and \texttt{new}. The effect of the instruction  \texttt{y = CAS(T,x,old,new)}
  is conditional: the value
  of $x$ is replaced by \texttt{new} if it is equal to
  \texttt{old}, otherwise it does not change. The method returns
  \texttt{true} if the value changed, and \texttt{false} otherwise.

  A \texttt{CAS} instruction can be seen as a synchronization
  between two processes: $P_T$ associated with the thread $T$, and
  $P_x$ associated with the variable $x$. The states of $P_T$ are
  valuations of the local variables of $T$. The states of $P_x$ are
  the values $x$ can take. An instruction  of the form \texttt{y =
    CAS(T,x,old,new)} becomes a
  synchronization action between $P_T$ and $P_x$ with the
  two transitions of Figure~\ref{fig:cas} (represented for convenience as Petri net
  transitions).

  \begin{figure}
    \centering
     \begin{minipage}{.49\linewidth}
 \begin{center}
     \begin{tikzpicture}[scale=.9,node distance=2cm,auto,>= triangle 45,bend angle = 30]
     \draw (0.6,0) node  {$s$};
\draw (.6,0) circle [radius=10pt];
\draw (.6,.6) node {$P_T$};
\draw (2,0) circle [radius=10pt]; 
\draw (2,.6) node {$P_x$};
\draw (2,0) node  {\texttt{old}};
\draw[thick] (1,-1.5) -- (1.6,-1.5);
\draw (.6,-.25) node (s1) {}; \draw (2cm,-.25) node (s2) {}; \draw (1.3,-1.5) node (m) {}; 
\draw (3.5,-1.5) node {\texttt{y = CAS(T,x,old,new)}};
\draw[->] (s1) -- (m);
\draw[->] (s2) -- (m);
\draw (.6,-3) circle [radius=10pt];
\draw (2,-3) circle [radius=10pt]; 
\draw (.6,-2.75) node (t1) {}; \draw (2,-2.75) node (t2) {};
\draw[->] (m) -- (t1);
\draw[->] (m) -- (t2);
\draw (0.6,-3) node {$s'$};
\draw (2,-3) node {\texttt{new}};
   \end{tikzpicture}
  \end{center}
   \end{minipage}
   \begin{minipage}{.49\linewidth}
 \begin{center}
     \begin{tikzpicture}[scale=.9,node distance=2cm,auto,>= triangle 45,bend angle = 30]
     \draw (0.6,0) node  {$s$};
\draw (.6,0) circle [radius=10pt];
\draw (.6,.6) node {$P_T$};
\draw (2,0) circle [radius=10pt]; 
\draw (2,.6) node {$P_x$};
\draw (2,0) node  {\texttt{v}};
\draw[thick] (1,-1.5) -- (1.6,-1.5);
\draw (.6,-.25) node (s1) {}; \draw (2cm,-.25) node (s2) {}; \draw
(1.3,-1.5) node (m) {}; 
\draw (3.5,-1.5) node {\texttt{y = CAS(T,x,old,new)}};
%\draw (2,-1.5) node {$a$};
\draw[->] (s1) -- (m);
\draw[->] (s2) -- (m);
\draw (.6,-3) circle [radius=10pt];
\draw (2,-3) circle [radius=10pt]; 
\draw (.6,-2.75) node (t1) {}; \draw (2,-2.75) node (t2) {};
\draw[->] (m) -- (t1);
\draw[->] (m) -- (t2);
\draw (0.6,-3) node {$s''$};
\draw (2,-3) node {\texttt{v}};
   \end{tikzpicture}
  \end{center}
   \end{minipage}  
    \caption{\texttt{CAS} as transitions of a Zielonka automaton. On the left side of the figure we have the case when the value of $x$ is
   \texttt{old}; on the right side \texttt{v} is different from \texttt{old}. Notice that in
   state $s'$ the value of $y$ is \texttt{true}, whereas in $s''$, it is \texttt{false}.}
    \label{fig:cas}
  \end{figure}
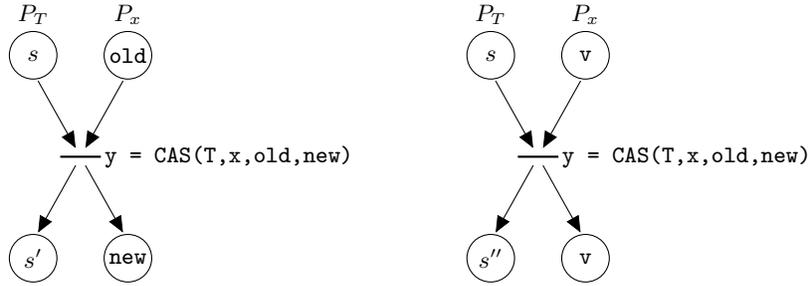
\end{example}

A Zielonka automaton can be seen as a usual finite-state
automaton, whose set of states  $S = \prod_{p\in\PP} S_p$ is given by the
global states, and transitions $s \act{a} s'$ if $(s_{\loc(a)},a,
s'_{\loc(a)}) \in \d$, and $s_{\PP\setminus
  \loc(a)}=s'_{\PP\setminus \loc(a)}$. Thus states of this automaton
are tuples of states of the processes of the Zielonka
automaton. As a language acceptor, a Zielonka automaton $\Aa$ accepts
a \emph{trace-closed language} $L(\Aa)$, that is, a language closed
under commuting adjacent independent
symbols. Formally, a language $L$ is trace-closed when $uabv \in
L$ if and only if  $ubav \in L$, for all $u,v \in \S^*$ and all 
independent actions $a,b$.

A cornerstone result in the theory of Mazurkiewicz traces is a
construction transforming a sequential automaton into an equivalent
deterministic Zielonka automaton.   This beautiful result is one of the
rare examples of distributed synthesis with broader scope.

\begin{theorem} \label{th:zielonka} \cite{zie87} Given a distributed alphabet
  $(\S,\loc)$, and a regular trace-closed language $L \subseteq \S^*$
  over $(\S,\loc)$. A
  deterministic Zielonka automaton $\Aa$ such that $L(\Aa)=L$ can be
  effectively constructed.
\end{theorem}

The only assumption of the theorem above is that the language of the
automaton is trace-closed, but this is unavoidable.  Moreover, trace
closure can be checked easily, e.g.~on the minimal DFA of the given
language.

The construction behind Theorem~\ref{th:zielonka} is technically
involved, but also very fascinating. The crux is to show
how to put together distributed information using additional memory
that is \emph{finite}\footnote{Vector clocks~\cite{matt88} are a
  similar notion in distributed computing, but they do not require a
  finite domain of values.}.  Many researchers contributed to simplify the construction and
to improve its complexity, see~\cite{cmz93,ms97,GM06,GGMW10}
and references therein. The most recent construction~\cite{GGMW10} produces
deterministic Zielonka automata of size that is exponential in the
number of processes (and polynomial in the size of a DFA for $L$). The
exponential dependence on the number of processes is
necessary, modulo a technical assumption (that is actually required for
monitoring).

\begin{theorem}[\cite{GGMW10}]
  There is an algorithm that takes as input a distributed alphabet
  $(\S,\loc)$ over $n$ processes and a DFA $\Aa$ accepting a trace-closed
  language over $(\S,\loc)$, and computes an equivalent deterministic
  Zielonka automaton $\Bb$ with at most $4^{n^4} \cdot |\Aa|^{n^2}$ states
  per process. Moreover, the algorithm computes the transitions of
  $\Bb$ on-the-fly in polynomial time. % , and checks whether a state is
%   final in polynomial time as well.
\end{theorem}

\section{Distributed monitoring}\label{sec:moni}

The construction of deterministic Zielonka automata opens interesting
perspectives for monitoring concurrent programs. In order to monitor
a concurrent program at runtime, the monitor has to be distributed (or
decentralized). This means that there is a local monitor on each
thread, and these local monitors can exchange information. The
exchange can be implemented by allowing local monitors to initiate extra
communication, or, more conservatively, by using the available
communication in the program in order to share monitoring-relevant
information. We follow the latter setting here, since adding
communication can reduce the concurrency, and  it is very difficult to
quantify how much performance is lost by adding communication.

Apart from detecting violations of safety properties at runtime, the
information gathered by such monitors can be also used to
recover from an unsafe state. Of course, this can be done only at
runtime, and not offline, by inspecting sequential executions a posteriori.

Our general approach for this kind of distributed monitoring is
simple: we have some 
trace-closed, regular property $\phi$ that should be satisfied by every
execution of a given program or system. To detect possible
violations of $\phi$  at runtime, we construct a monitor for $\phi$ and
run it in
parallel with the program. Consider the scenario where the program $P$
is modeled by a Zielonka
automaton $\Aa_P$. If a monitor is also a Zielonka automaton $\Aa_M$,
then running the monitor $M$ in parallel to $P$ amounts to build the usual product automaton between
$\Aa_P$ and $\Aa_M$ process-wise.

Interestingly, many properties one is interested to monitor on
concurrent programs can be  expressed in terms of the
happens-before relation between specific events, as the following example illustrates.

\begin{example}
Consider the \emph{race detection problem} from
Section~\ref{ssec:races}. A race occurs when two
conflicting accesses to the same variable are unordered in the happens-before
relation. Therefore, a violation of the ``no-race''
property is monitored by looking for two \emph{unordered} 
accesses to the same variable, at least one of them being a write. 

Monitoring a violation of \emph{atomicity} (recall
Section~\ref{ssec:atomic}) is done by 
checking for every transaction on some thread $T$, that no action $c$ of some
thread $T' \not= T$ happened after the beginning $a=\bg(T)$ of the
transaction  on $T$ (cf.~instruction 1 of Example~\ref{ex:atomic}) and before its matching end
$b=\en(T)$ (cf.~instruction 4). In other
words, the monitor looks for events $c$ on $T' \not= T$ satisfying $a
\prec c \prec b$ in the happens-before relation.
\end{example}

Determining the partial ordering between specific events is 
closely related to the kernel of all available constructions
behind Zielonka's theorem. This is known as
the~\emph{gossip automaton}~\cite{ms97}, and the name reflects its
r\^ole: it computes what a process knows about the knowledge of other
processes. Using \emph{finite-state} gossiping, processes can put
together information that is distributed in the system, hence reconstruct the
execution of the given DFA.

The gossip automaton is already responsible for the exponential
complexity of Zielonka automata, in all available constructions. A natural question is whether the construction of the gossip
automaton can be avoided, or at least simplified. Perhaps
unsurprisingly, the theorem below shows that gossiping is not needed
when the communication structure is acyclic.

The \emph{communication graph}  of a distributed alphabet
$(\S,\loc)$  with unary or binary action domains is the undirected
graph where vertices are the processes, and edges relate processes $p
\not= q$ if
there is some action $a \in \S$ such that $\loc(a)=\set{p,q}$.

\begin{theorem}[\cite{km13tcs}]
  Let $(\S,\loc)$ be a distributed alphabet with acyclic communication
  graph. Every regular, 
  trace-closed language $L$ over $\S$ can be accepted by a deterministic
  Zielonka automaton with $O(s^2)$ states per process, where $s$ is
  the size of the minimal DFA for $L$.
\end{theorem}

The theorem above can be useful to monitor programs with acyclic
communication if we can start from a small DFA for the trace-closed
language $L$ representing the monitoring property. However, in some
cases the DFA is necessarily large because it needs to take into
account many interleavings. For example, monitoring for some unordered
occurrences of $b$ and $c$, requires a DFA to remember \emph{sets} of
actions. In this case it is more efficient to start with a description
of $L$ by partial orders. We discuss a solution for this setting in
Section~\ref{ssec:gossip} below.

We need to emphasize that  using Zielonka automata for
monitoring properties in practice does not depend only on the efficiency of
the constructions from the above theorems. In addition to determinism,
further properties are desirable when building monitoring automata.
The first requirement is that a violation of the property to monitor
should be detectable locally. The reason for
this is that a thread that knows about the failure can start some
recovery actions or inform other threads about the failure. Zielonka
automata satisfying this property are called \emph{locally
  rejecting}~\cite{GGMW10}. More formally, each process $p$ has a
subset of states $R_p \subseteq S_p$; an execution leads a process $p$
into a state from $R_p$ if and only if $p$ knows already that the
execution cannot be extended to a trace in $L(\Aa)$.  The second
important requirement is that the monitoring automaton $\Aa_M$ should
not block the monitored system. This can be achieved by asking that in
every global state of $\Aa_M$ such that no process is in a rejecting
state, every action is enabled. A related discussion on desirable
properties of Zielonka automata and on  implementating the
construction of \cite{GGMW10} is reported in \cite{adgs13}.

\subsection{Gossip in trees}\label{ssec:gossip}

In this section we describe a setting where we can compute efficiently
the happens-before relation for selected actions of a concurrent
program. We will first illustrate  the idea on the simple example of
Section~\ref{ssec:atomic}. The program there has two threads, $T_1$
and $T_2$, and one shared variable $x$. For clarity we add actions
$\bg(T_i),\en(T_i)$ denoting the begin/end of the
atomic section on $T_i$. Thus, the alphabet of actions is:
\[
\S=\set{\bg(T_1),\en(T_1),w(T_1,x),r(T_1,x), \bg(T_2),\en(T_2),w(T_2,x)}\,.
\] 
The dependence relation $D$ includes all
pairs of actions of the same thread, as well as the pairs $(r(T_1,x),w(T_2,x))$
and $(w(T_1,x),w(T_2,x))$. Following Example~\ref{ex:procs}, the
Zielonka automaton has processes $\PP=\set{T_1,T_2,\struct{T_1,x},\struct{T_2,x}}$
and the domains of actions are:
\begin{eqnarray*}
 && \loc(\bg(T_i))=\loc(\en(T_i))=\set{T_i},\\
&& \loc(r(T_i,x))=\set{T_i,\struct{T_i,x}}\\
&& \loc(w(T_i,x))=\set{T_i,\struct{T_i,x},\struct{T_{3-i},x}}
\end{eqnarray*}

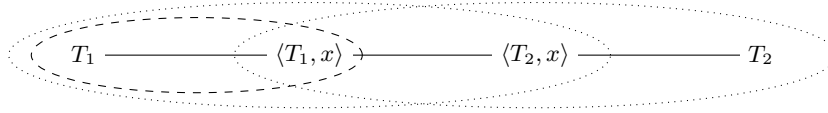
\begin{figure}
  \centering
  \begin{tikzpicture}
  \draw (0,0) node [fill=white] {$T_1$} -- 
(3,0) node [fill=white]   {$\struct{T_1,x}$} --
(6,0) node [fill=white]   {$\struct{T_2,x}$} -- 
(9,0) node  [fill=white]  {$T_2$};
\draw[dashed] (1.5,0) ellipse [x radius=2.2cm,y radius=.5cm];
%\draw[dashed] (7.5,0) ellipse [x radius=2.2cm,y radius=.5cm];
\draw[dotted] (3,0) ellipse [x radius=4cm,y radius=.7cm];
\draw[dotted] (6,0) ellipse [x radius=4cm,y radius=.7cm];
\end{tikzpicture}
  \caption{A tree of processes $\Tt$. The dashed part is
the domain of the read action, whereas the dotted parts are the domains of
the two writes.}
  \label{fig:ptree}
\end{figure}

Note that we can represent these four processes as a (line) tree in which
the domain of each action spans a connected part of the tree, see
Figure~\ref{fig:ptree}. 

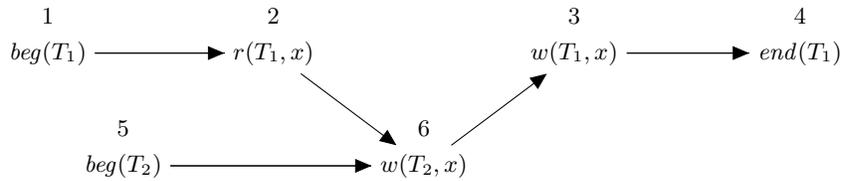
\begin{figure}
  \centering
  \begin{tikzpicture}[>= triangle 45]
\draw (0,.5) node {1}; \draw (3,.5) node {2};
\draw (0,0) node[fill=white] (n1) {$\bg(T_1)$} -- (3,0) node
[fill=white] (n2) {$r(T_1,x)$} ;
\draw[->] (n1) -- (n2);
\draw (1,-1) node {5}; \draw (5,-1) node {6};
\draw (1,-1.5) node[fill=white] (n5) {$\bg(T_2)$} -- (5,-1.5) node
[fill=white] (n6) {$w(T_2,x)$} ;
\draw[->] (n5) -- (n6);
\draw[->] (n2) -- (n6);
\draw (7,.5) node {3}; \draw (10,.5) node {4};
\draw (7,0) node[fill=white] (n3) {$w(T_1,x)$} -- (10,0) node
[fill=white] (n4) {$\en(T_1)$} ;
\draw[->] (n6) -- (n3);
\draw[->] (n3) -- (n4);
\end{tikzpicture}
  \caption{Violation of conflict serializability (partial order).}
  \label{fig:trace}
\end{figure}

\begin{figure}[h]
\phantom{bb} \hspace{-.5cm} 
  %\centering
\begin{tikzpicture}[scale=.9]
  \draw (0.5,0) node {$T_1$} (0.5,-2) node {$\struct{T_1,x}$} (0.5,-4) node
  {$\struct{T_2,x}$} (0.5,-6) node {$T_2$};
\draw (2,1.2) node {1} (4.5,1.2) node {2} (7,1.2) node {5} (12,1.2) node {6};
\draw (2,0) node {$\bg(T_1)$} (2,-2) node {$\es$} (2,-4) node {$\es$}
(2,-6) node {$\es$}; 
\draw[rounded corners] (1.3,-.5) rectangle (2.7,.5);

\draw (3.8,0) node (n1a) {$\bg(T_1)$} (3.8,-2) node (n1b) {$\bg(T_1)$} (4.5,-4) node
{$\cdots$} (4.5,-6) node {$\cdots$}; 
\draw (5.4,0) node (n2a) {$r(T_1,x)$} (5.4,-2) node (n2b) {$r(T_1,x)$};
\draw[->] (n1a) -- (n2a);
\draw[->] (n1b) -- (n2b);
\draw[rounded corners] (3.1,-.5) rectangle (6.1,.5);
\draw[rounded corners,thick] (3.1,-2.5) rectangle (6.1,-1.5);

\draw (7,0) node {$\cdots$} (7,-2) node {$\cdots$} (7,-4) node {$\cdots$}
(7,-6) node {$\bg(T_2)$}; 
\draw[rounded corners,thick] (6.3,-6.4) rectangle (7.7,-5.6);

\draw (10.5,-6.5)  node (n4a) {$\bg(T_2)$} (10.5,-4.5)  node (n4b)
{$\bg(T_2)$} (10.5,-2.5) node (n4c) {$\bg(T_2)$};
\draw (13,-6.5) node (n5a) {$w(T_2,x)$} (13,-4.5) node (n5b) {$w(T_2,x)$}
(13,-2.5) node (n5c) {$w(T_2,x)$};
\draw (11.5,-5.5) node (n2c) {$r(T_1,x)$} (11.5,-3.5)  node (n2d){$r(T_1,x)$}
(11.5,-1.5)  node (n2e) {$r(T_1,x)$} ;
\draw (9.5,-5.5) node (n1c) {$\bg(T_1)$} (9.5,-3.5)  node (n1d) {$\bg(T_1)$}
(9.5,-1.5)  node (n1e)  {$\bg(T_1)$};
\draw[->] (n4a) -- (n5a);
\draw[->] (n4b) -- (n5b);
\draw[->] (n4c) -- (n5c);

\draw[->] (n2c) -- (n5a);
\draw[->] (n2d) -- (n5b);
\draw[->] (n2e) -- (n5c);

\draw[->] (n1c) -- (n2c);
\draw[->] (n1d) -- (n2d);
\draw[->] (n1e) -- (n2e);
\draw[rounded corners] (8.7,-2.8) rectangle (14,-1.2);
\draw[rounded corners] (8.7,-4.8) rectangle (14,-3.2);
\draw[rounded corners] (8.7,-6.8) rectangle (14,-5.2);

\draw (12,0) node {$\cdots$};
\end{tikzpicture}  

  \caption{Computing the partial order. Numbers in the first line
    stand for the program lines in Example~\ref{ex:atomic}. Dots mean that the information does
    not change.}
  \label{fig:gossip}
\end{figure}
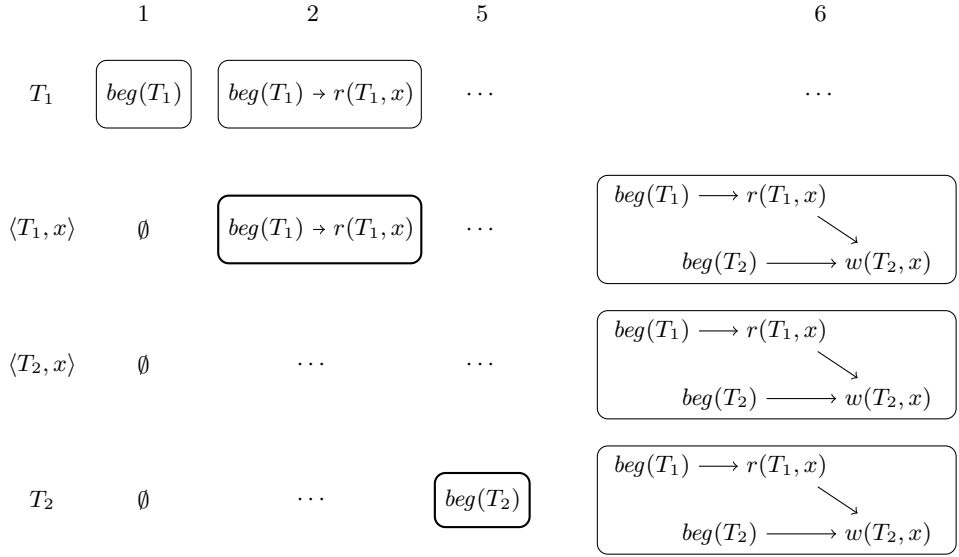

The Mazurkiewicz trace in Figure~\ref{fig:trace}, represented as a
partial order, shows a violation of conflict serializability: event $c$
at step~6 satisfies $a \prec c \prec b$, where $a$ represents step~1, and
$b$ step~4. The happens-before relation
can be computed piecewise by a Zielonka automaton, by
exchanging information via the synchronization
actions. Figure~\ref{fig:gossip} illustrates how processes update
their knowledge about the partial order. Note how the two  partial
orders represented by thick lines, are combined together with the
action $w(T_2,x)$ of step 6, in order to produce the partial orders of processes
$\struct{T_1,x}$, $\struct{T_2,x}$ and $T_2$ in the last column. Thus,
after step 6 these processes know that action $w(T_2,x)$ happened
after $\bg(T_1)$. Executing then steps 3 and 4 will inform process
$T_1$ about the violation of conflict serializability.

Gossiping in trees of processes works more generally as follows. We call
a distributed alphabet $(\S,\loc)$ over process set $\PP$ 
\emph{tree-like}, if there exists some tree $\Tt$ with $\PP$ as set of
vertices, such that the domain of each
action is a \emph{connected} part of $\Tt$. 

Note that the tree $\Tt$ is uniquely defined by action domains in the
special case where the actions have at most binary domains. Otherwise,
there can be several trees as above, but we will assume that the
distributed alphabet comes with a suitable tree representation.

We are also given a set of monitoring actions $\G \subseteq \S$. The
task is to compute for each process $p \in\PP$ the
\emph{happens-before relation w.r.t.~$\G$}, in other words the
happens-before relation between the most recent occurrences of actions
from $\G$ that are known to process $p$. This information is a DAG where the
set of nodes is a subset of $\G$. Figure~\ref{fig:gossip} below gives
an example of such a computation.

\begin{theorem}\label{th:gossip}
  Given a tree-like distributed alphabet $(\S,\loc)$ with tree $\Tt$, and a
  set $\G \subseteq \S$ of actions. The happens-before relation
  w.r.t.~$\G$ can be computed by a Zielonka automaton where every
  process $p$ maintains two DAGs of size $|\G|+\out(p)$, with $\out(p)$
  the out-degree of $p$ in $\Tt$. Each update of the DAGs can be
  done in linear time in their size.
\end{theorem}

Theorem~\ref{th:gossip} provides a rather simple way of reconstructing
the happens-before relation with \emph{finite} additional memory, and
in a distributed way. Each synchronization action $b$ will update the
DAGs maintained by the processes executing $b$, by combining these
DAGs and selecting the most recent knowledge about actions of $\G$. As
an example, suppose that processes $p$ and $q$ are neighbors in the
tree, say, $q$ is the father of $p$. As long as there  is no
synchronization involving both $p$ and 
$q$, process $p$ has the most recent knowledge about occurrences of
$\G$-actions belonging to processes in the subtree of $p$. As soon
as some action synchronizes $p$ and $q$, process $q$ will be able to
include $p$'s knowledge regarding these actions, in its own knowledge.

%\section{Distributed Control}

\section{Related work}

This brief overview aimed at presenting the motivation behind
distributed synthesis and how Mazurkiewicz trace theory can be used
for monitoring concurrent programs. To conclude we point out some
further related results.

Our discussion turned around distributed synthesis in a simple
case where the program evolves without external actions from an
environment. Synthesis of open systems, i.e., systems with an
environment, is a more complex problem. Synthesis started as a
problem in logics, with Church's problem asking for an algorithm to
construct devices that transform sequences of input bits into
sequences of output bits, so that the interaction conforms to a given logical
formula~\cite{church62}. Later, Ramadge and Wonham proposed the
\emph{supervisory control} formulation~\cite{RW89}, where a plant and
a specification are given; a controller should be designed such that
its product with the plant satisfies the specification, no matter what
the environment does.
Synthesis is a particular case of control where the plant allows for
every possible behavior. Rabin's result on the decidability of monadic
second-order logic over infinite trees answered Church's question for
MSO specifications~\cite{rab72}.

\paragraph{Synthesis without environment.} 
 The problem of distributed synthesis without environment was first raised in the context of
 Petri nets. The task there is to decide whether an automaton, viewed
 as a graph, is isomorphic to the marking graph of a net. 
Ehrenfeucht and Rozenberg introduced the notion of
 regions, that determines how to decompose a graph for obtaining a
 net~\cite{ER89}.

Zielonka's algorithm has
been applied to solve the synthesis problem for models that go beyond
Mazurkiewicz traces. One example is synthesis of communicating
automata from graphical specifications known as \emph{message sequence
  charts}. Communicating automata are distributed finite-state
automata communicating over point-to-point FIFO channels. As such, the
model is Turing powerful. However, if the communication channels are
bounded, there is a tight link between execution sequences of the
communicating automaton and Mazurkiewicz
traces~\cite{HMNST03}. Actually we can handle even the case where the
assumption about bounded channels is relaxed by asking that they 
are bounded for \emph{at least one} scheduling of message
receptions~\cite{GKM06}. Producer-consumer behaviors are captured by
this relaxed requirement.

Multiply nested words with
various kinds of bounds on 
stack access~\cite{qr05,LMP07,LP12}, are an attractive 
model for concurrent programs with recursion, because of decidability
properties 
and expressiveness. In \cite{BGH09} the model is extended to nested
Mazurkiewicz traces and Zielonka's construction  is lifted to this
setting.

For runtime monitoring, a  similar approach as ours is advocated
in~\cite{svar06}, that proposes  an epistemic temporal logic for
describing safety properties. A distributed implementation of a
monitor for such properties  is obtained,
based on a variant of vector clocks.

\paragraph{Synthesis with environment.} 

One way to lift Church's problem to the distributed setting was
proposed by Pnueli and
Rosner~\cite{PR90}. They showed that synthesis is
decidable for very restricted architectures, namely
pipelines. General undecidability was already known from the
framework on multi-player
games~\cite{PetRei79}. Subsequently, \cite{MadThiag01} showed that pipelines are
essentially the only decidable case. Some interesting extensions of
Pnueli and Rosner's setting have been considered in~\cite{KV01,FinSch05,gs13tocl}.

Alternatively, a distributed formulation of
Church's problem can be formulated in Ramadge
and Wonham's supervisory control setting. This problem, when plants
and controllers are
Zielonka automata, has been considered in~\cite{GLZ04,MTY05,GGMW13,MW14}. In this formulation,
local controllers exchange information when synchronizing on shared actions.
In this way, the arguments for undecidability based on hidden information,
as in~\cite{PR90,PetRei79}, do not apply.
Decidability of distributed control for Zielonka automata is still
open. It is known though that this formulation admits more decidable
architectures: the control problem for local parity specifications is decidable over acyclic
architectures~\cite{MW14}, thus in cases where Pnueli and
Rosner's model is undecidable.

Yet another distributed version of Ramadge and Wonham's problem is
considered in~\cite{dr12}, this time for Petri nets. The problem is to
compute local controllers that guarantee basic properties like
e.g.~deadlock avoidance. The limitation of this approach is that the
algorithm may fail to find local controllers, although they 
exist. 

Game semantics and asynchronous games played on event structures are
introduced in~\cite{mel06}. Such games are investigated
in~\cite{gutwin14} from a game-theoretical viewpoint, showing a Borel
determinacy result under some restrictions.

\paragraph{Verification.}
As we already mentioned, automated verification of concurrent systems
encounters major problems due to state explosion. One
particularly efficient technique able to addresses this problem is
known as \emph{partial order reduction} (POR)
\cite{gp93,pel93,val91}. It consists of restricting the exploration of
the state space by avoiding the execution of similar, or equivalent
runs. The notion of equivalence of runs used by POR is based on Mazurkiewicz traces. The efficiency of POR
methods depends of course on the precise equivalence notion between
executions. Recent variants, such as dynamic POR, work without
storing explored states explicitly and aim at improving the precision
by computing additional information about (non)-equivalent
executions~\cite{dynPORpopl14}.

There are other contexts in verification where analysis gets more
efficient using equivalences based on Mazurkiewicz traces. One such
setting is counter-example generation based on partial (Mazurkiewicz)
traces instead of linear executions
\cite{cerny+cav13}. Previous work connecting concurrency issues and
Mazurkiewicz trace theory concerns atomicity
violations~\cite{fm08cav}, non-linearizability and sequential
inconsistency~\cite{amp96lics}.

\medskip

\emph{Acknowledgments.} Very special thanks to 
J\'er\^ome Leroux, Gabriele Puppis and  Igor Walukiewicz for numerous
comments on previous versions of this paper.

\bibliographystyle{abbrv}
\bibliography{distrib,DistrGames}

\begin{thebibliography}{10}

\bibitem{dynPORpopl14}
P.~Abdulla, S.~Aronis, B.~Jonsson, and K.~Sagonas.
\newblock Optimal dynamic partial order reduction.
\newblock In {\em POPL'14}, pages 373--384. ACM, 2014.

\bibitem{adgs13}
S.~Akshay, I.~Dinca, B.~Genest, and A.~Stefanescu.
\newblock Implementing realistic asynchronous automata.
\newblock In {\em FSTTCS'13}, LIPIcs, pages 213--224. Schloss Dagstuhl -
  Leibniz-Zentrum fuer Informatik, 2013.

\bibitem{amp96lics}
R.~Alur, K.~McMillan, and D.~Peled.
\newblock Model-checking of correctness conditions for concurrent objects.
\newblock In {\em LICS'96}, pages 219--228. IEEE, 1996.

\bibitem{erlang07}
J.~Armstrong.
\newblock {\em Programming Erlang: Software for a Concurrent World}.
\newblock Pragmatic Bookshelf, 2007.

\bibitem{BGH09}
B.~Bollig, M.-L. Grindei, and P.~Habermehl.
\newblock Realizability of concurrent recursive programs.
\newblock In {\em FoSSaCS'09}, LNCS, pages 410--424. Springer, 2009.

\bibitem{cerny+cav13}
P.~Cern{\'y}, T.~A. Henzinger, A.~Radhakrishna, L.~Ryzhyk, and T.~Tarrach.
\newblock Efficient synthesis for concurrency by semantics-preserving
  transformations.
\newblock In {\em CAV'13}, LNCS, pages 951--967. Springer, 2013.

\bibitem{church62}
A.~Church.
\newblock Logic, arithmetics, and automata.
\newblock {\em Proceedings of the International Congress of Mathematicians},
  1962.

\bibitem{cmz93}
R.~Cori, Y.~M{\'e}tivier, and W.~Zielonka.
\newblock Asynchronous mappings and asynchronous cellular automata.
\newblock {\em Information and Computation}, 106:159--202, 1993.

\bibitem{dr12}
P.~Darondeau and L.~Ricker.
\newblock Distributed control of discrete-event systems: A first step.
\newblock {\em Transactions on Petri Nets and Other Models of Concurrency},
  6:24--45, 2012.

\bibitem{DR95}
V.~Diekert and G.~Rozenberg, editors.
\newblock {\em {The Book of Traces}}.
\newblock World Scientific, Singapore, 1995.

\bibitem{ER89}
A.~Ehrenfeucht and G.~Rozenberg.
\newblock Partial (set) 2-structures: Parts i and ii.
\newblock {\em Acta Informatica}, 27(4):315--368, 1989.

\bibitem{fm08cav}
A.~Farzan and P.~Madhusudan.
\newblock Monitoring atomicity in concurrent programs.
\newblock In {\em CAV'08}, LNCS, pages 52--65. Springer, 2008.

\bibitem{FinSch05}
B.~Finkbeiner and S.~Schewe.
\newblock Uniform distributed synthesis.
\newblock In {\em LICS'05}, pages 321--330. IEEE, 2005.

\bibitem{GLZ04}
P.~Gastin, B.~Lerman, and M.~Zeitoun.
\newblock Distributed games with causal memory are decidable for
  series-parallel systems.
\newblock In {\em FSTTCS'04}, LNCS, pages 275--286. Springer, 2004.

\bibitem{gs13tocl}
P.~Gastin and N.~Sznajder.
\newblock Fair synthesis for asynchronous distributed systems.
\newblock {\em ACM Transactions on Computational Logic}, 14(2):9, 2013.

\bibitem{GGMW10}
B.~Genest, H.~Gimbert, A.~Muscholl, and I.~Walukiewicz.
\newblock Optimal {Z}ielonka-type construction of deterministic asynchronous
  automata.
\newblock In {\em ICALP'10}, LNCS, pages 52--63. Springer, 2010.

\bibitem{GGMW13}
B.~Genest, H.~Gimbert, A.~Muscholl, and I.~Walukiewicz.
\newblock Asynchronous games over tree architectures.
\newblock In {\em ICALP'13}, LNCS, pages 275--286. Springer, 2013.

\bibitem{GKM06}
B.~Genest, D.~Kuske, and A.~Muscholl.
\newblock A {K}leene theorem and model checking algorithms for existentially
  bounded communicating automata.
\newblock {\em Inf. Comput.}, 204(6):920--956, 2006.

\bibitem{GM06}
B.~Genest and A.~Muscholl.
\newblock {Constructing Exponential-Size Deterministic Zielonka Automata}.
\newblock In {\em ICALP'06}, LNCS, pages 565--576. Springer, 2006.

\bibitem{gp93}
P.~Godefroid and P.~Wolper.
\newblock Using partial orders for the efficient verification of deadlock
  freedom and safety properties.
\newblock {\em Formal Methods in System Design}, 2(2):149--164, 1993.

\bibitem{gutwin14}
J.~Gutierrez and G.~Winskel.
\newblock On the determinacy of concurrent games on event structures with
  infinite winning sets.
\newblock {\em J. Comput. Syst. Sci.}, 80(6):1119--1137, 2014.

\bibitem{HMNST03}
J.~G. Henriksen, M.~Mukund, K.~N. Kumar, M.~Sohoni, and P.~S. Thiagarajan.
\newblock {A Theory of Regular {MSC} Languages}.
\newblock {\em Inf. Comput.}, 202(1):1--38, 2005.

\bibitem{km13tcs}
S.~Krishna and A.~Muscholl.
\newblock A quadratic construction for {Z}ielonka automata with acyclic
  communication structure.
\newblock {\em Theoretical Computer Science}, 503:109--114, 2013.

\bibitem{KV01}
O.~Kupferman and M.~Y. Vardi.
\newblock Synthesizing distributed systems.
\newblock In {\em LICS'01}, pages 389--398. IEEE, 2001.

\bibitem{LMP07}
S.~{La Torre}, P.~Madhusudan, and G.~Parlato.
\newblock A robust class of context-sensitive languages.
\newblock In {\em LICS'07}, pages 161--170. IEEE, 2007.

\bibitem{LP12}
S.~{La Torre} and G.~Parlato.
\newblock Scope-bounded multistack pushdown systems: fixed-point,
  sequentialization, and tree-width.
\newblock In {\em FSTTCS'12}, LIPIcs, pages 173--184. Schloss Dagstuhl -
  Leibniz-Zentrum fuer Informatik, 2012.

\bibitem{lamport78}
L.~Lamport.
\newblock Time, clocks, and the ordering of events in a distributed system.
\newblock {\em Operating Systems}, 21(7):558--565, 1978.

\bibitem{MadThiag01}
P.~Madhusudan and P.~Thiagarajan.
\newblock Distributed control and synthesis for local specifications.
\newblock In {\em ICALP'01}, volume 2076 of {\em LNCS}, pages 396--407.
  Springer, 2001.

\bibitem{MTY05}
P.~Madhusudan, P.~S. Thiagarajan, and S.~Yang.
\newblock The {MSO} theory of connectedly communicating processes.
\newblock In {\em FSTTCS'05}, LNCS, pages 201--212. Springer, 2005.

\bibitem{matt88}
F.~Mattern.
\newblock Virtual time and global states of distributed systems.
\newblock In {\em International Workshop on Parallel and Distributed
  Algorithms}, pages 215--226. Elsevier, 1989.

\bibitem{maz77}
A.~Mazurkiewicz.
\newblock Concurrent program schemes and their interpretations.
\newblock {DAIMI Rep. PB}~78, Aarhus University, Aarhus, 1977.

\bibitem{mel06}
P.-A. Melli\`es.
\newblock Asynchronous games 2: {T}he true concurrency of innocence.
\newblock {\em TCS}, 358(2-3):200--228, 2006.

\bibitem{ms97}
M.~Mukund and M.~A. Sohoni.
\newblock Keeping track of the latest gossip in a distributed system.
\newblock {\em Distributed Computing}, 10(3):137--148, 1997.

\bibitem{MW14}
A.~Muscholl and I.~Walukiewicz.
\newblock Distributed synthesis for acyclic architectures.
\newblock In {\em FSTTCS'14}, LIPIcs, pages 639--651. Schloss Dagstuhl -
  Leibniz-Zentrum fuer Informatik, 2014.

\bibitem{scala14}
M.~Odersky and T.~Rompf.
\newblock Unifying functional and object-oriented programming with {Scala}.
\newblock {\em Communications of the ACM}, 57(4):76--86, 2014.

\bibitem{scala-book}
M.~Odersky, L.~Spoon, and B.~Venners.
\newblock {\em Programming in {Scala}}.
\newblock Artima, 2010.

\bibitem{pel93}
D.~A. Peled.
\newblock All from one, one for all: on model checking using representatives.
\newblock In {\em CAV'93}, LNCS, pages 409--423. Springer, 1993.

\bibitem{PetRei79}
G.~L. Peterson and J.~H. Reif.
\newblock Multi-person alternation.
\newblock In {\em FOCS'79}, pages 348--363. IEEE, 1979.

\bibitem{PR90}
A.~Pnueli and R.~Rosner.
\newblock Distributed reactive systems are hard to synthesize.
\newblock In {\em FOCS'90}. IEEE, 1990.

\bibitem{qr05}
S.~Qadeer and J.~Rehof.
\newblock Context-bounded model-checking of concurrent software.
\newblock In {\em TACAS'05}, LNCS, pages 93--107. Springer, 2005.

\bibitem{rab72}
M.~O. Rabin.
\newblock {\em Automata on Infinite Objects and Church's Problem}.
\newblock American Mathematical Society, Providence, RI, 1972.

\bibitem{RW89}
P.~Ramadge and W.~Wonham.
\newblock The control of discrete event systems.
\newblock {\em Proceedings of the IEEE}, 77(2):81--98, 1989.

\bibitem{svar06}
K.~Sen, A.~Vardhan, G.~Agha, and G.~Rosu.
\newblock Decentralized runtime analysis of multithreaded applications.
\newblock In {\em International Parallel and Distributed Processing Symposium
  (IPDPS 2006)}. IEEE, 2006.

\bibitem{val91}
A.~Valmari.
\newblock Stubborn sets for reduced state space generation.
\newblock In {\em Advances in {P}etri {N}ets 1990}, number 483 in LNCS, pages
  491--515, 1991.

\bibitem{zie87}
W.~Zielonka.
\newblock Notes on finite asynchronous automata.
\newblock {\em RAIRO--Theoretical Informatics and Applications}, 21:99--135,
  1987.

\end{thebibliography}

\end{document}